\documentclass[longauth,letter,traditabstract]{aa}  
\usepackage{graphicx}
\usepackage{natbib}
\usepackage{amsmath}

\newcommand{\sub}[1]{_{\rm #1}}

\newcommand{\CII}{[C\,{\sc ii}]}
\newcommand{\CI}{[C\,{\sc i}]}
\newcommand{\HII}{H\,{\sc ii}}

\newcommand{\OI}{[O\,{\sc i}]}

\newcommand{\micron}{\,$\mu$m}

\usepackage{txfonts}
\begin{document}
\title{The origin of the \CII\/ emission in the S140 PDRs - new insights from HIFI \thanks{Herschel is an ESA space observatory with 
science instruments 
provided by European-led Principal Investigator consortia and with 
important participation from NASA.}}
\author{C.~Dedes\inst{1}$\!$, M.~R\"ollig\inst{2}, B.~Mookerjea\inst{3}, Y.~Okada\inst{2}, V.~Ossenkopf\inst{2,4}, S.~Bruderer\inst{1}, A.O.~Benz\inst{1}, M. Melchior\inst{5}, C.~Kramer\inst{6},
M.~Gerin\inst{7}, R.~G\"usten\inst{8},
M.~Akyilmaz\inst{2}, O.~Berne\inst{9},
F.~Boulanger\inst{10}, G.~De Lange\inst{4}, L.~Dubbeldam\inst{4}, K.~France\inst{11},
A.~Fuente\inst{12},  J.R.~Goicoechea \inst{13}, A.~Harris\inst{14}, R.~Huisman\inst{4}, W.~Jellema\inst{4}, 
C.~Joblin\inst{15,16}, T.~Klein\inst{8},  F.~Le Petit\inst{17},
S.~Lord\inst{18}, P.~Martin\inst{11}, J.~Martin-Pintado\inst{13},
 D.A.~Neufeld\inst{19}, S.~Philipp\inst{8}, T.~Phillips\inst{20}, P.~Pilleri\inst{15,16}, J.R.~Rizzo\inst{13}, M.~Salez\inst{21,7}, R.~Schieder\inst{2}, R.~Simon\inst{2}, O.~Siebertz\inst{2}, J.~Stutzki\inst{2},
F.~van der Tak\inst{4,22}, D.~Teyssier\inst{23}, H.~Yorke\inst{24}
}

\institute{Institute for Astronomy, ETH Z\"urich, 8093 Z\"urich, Switzerland
\and
I. Physikalisches Institut der Universit\"at 
zu K\"oln, Z\"ulpicher Stra\ss{}e 77, 50937 K\"oln, Germany
\and
Tata Institute of Fundamental Research (TIFR), Homi Bhabha Road, Mumbai 400005, India
\and
SRON Netherlands Institute for Space Research, P.O. Box 800, 9700 AV 
Groningen, Netherlands
\and
Institut fŸr 4D-Technologien, FHNW, 5210 Windisch, Switzerland
\and
Instituto de Radio Astronom\'ia Milim\'etrica (IRAM), Avenida Divina Pastora 7, Local 20, 18012 Granada, Spain
\and
LERMA, Observatoire de Paris, 61 Av. de l'Observatoire, 75014 Paris, France 
\and
Max-Planck-Institut f\"ur Radioastronomie, Auf dem H\"ugel 69, 53121, Bonn, Germany
\and
Leiden Observatory, Universiteit Leiden, P.O. Box 9513, NL-2300 RA Leiden, The Netherlands 
\and
Institut d'Astrophysique Spatiale, Universit\'e Paris-Sud, B\^at. 121, 91405 Orsay Cedex, France
\and
Department of Astronomy and Astrophysics, University of Toronto, 60 St. George Street, Toronto, ON M5S 3H8, Canada
\and
Observatorio Astron\'omico Nacional (OAN), Apdo. 112, 28803 Alcal\'a de Henares (Madrid), Spain
\and
Centro de Astrobiolog\'ia (INTA-CSIC), Ctra de Torrej\'on a Ajalvir, km 4, 28850 Torrej\'on de Ardoz, Madrid, Spain
\and
Astronomy Department, University of Maryland, College Park, MD 20742, USA
\and
Universit\'e de Toulouse, UPS, CESR, 9 avenue du colonel Roche, 31062 Toulouse cedex 4, France
\and
CNRS, UMR 5187, 31028 Toulouse, France
\and
Observatoire de Paris, LUTH and Universit\'e Denis Diderot, Place J. Janssen, 92190 Meudon, France
\and
IPAC/Caltech, MS 100-22, Pasadena, CA 91125, USA
\and 
Department of Physics and Astronomy, Johns Hopkins University, 3400 North Charles Street, Baltimore, MD 21218, USA
\and
California Institute of Technology, 320-47, Pasadena, CA  91125-4700, USA
\and
Laboratoire d'Etudes du Rayonnement et de la Mati\`ere en Astrophysique, UMR 8112 CNRS/INSU, OP, ENS, UPMC, UCP, Paris, France 
\and
Kapteyn Astronomical Institute, University of Groningen, PO box 800, 9700 AV Groningen, Netherlands
\and
European Space Astronomy Centre, Urb. Villafranca del Castillo, P.O. Box 50727, Madrid 28080, Spain
\and
Jet Propulsion Laboratory, 4800 Oak Grove Drive, MC 302-231, Pasadena, CA 91109  U.S.A.
}

\authorrunning {C.~Dedes\inst{1}  et al. } 
\titlerunning{The S140 PDR}

\abstract
{
Using {\em Herschel's} HIFI instrument we have observed \CII\/ along a
cut through S140 and high-$J$ transitions of CO and HCO$^+$
at  two positions on the cut,  corresponding to the externally
 irradiated ionization front and the embedded massive star
forming core IRS1. The HIFI data were combined with available
ground-based observations and modeled using the KOSMA-$\tau$ model for
photon dominated regions. Here we derive the physical conditions
in S140 and in particular the origin of \CII\/ emission around IRS1.
We identify three distinct regions of \CII\/ emission from the cut, one
close to the embedded source IRS1, one associated with the ionization
front and one further into the cloud.  The line emission can be
understood in terms of a clumpy model of photon-dominated regions. At
the position of IRS1, we identify at least two distinct components contributing to the \CII\/ emission, one of them a small, hot component, which can possibly be identified with the irradiated
outflow walls. This is consistent with the fact that the \CII\/ peak
at IRS1 coincides with shocked H$_2$ emission at the edges of the
outflow cavity.  We note that previously available observations of IRS1 can be well reproduced by a single-component KOSMA-$\tau$
model. Thus it is HIFI's unprecedented spatial and spectral
resolution, as well as its sensitivity which has allowed us to
uncover an additional hot gas component in the S140 region.}


\maketitle

\section{Introduction}

Massive star formation poses a challenge to the observer, since massive stars are formed predominantly in clustered environments and exhibit a number of entangled energetic phenomena in a small region. The S140 region shows several phenomena associated with (massive) star formation, such as outflows and strong UV irradiation both from internal and external heating sources creating photon dominated regions (PDRs). Thus, S140 could also act as a template for extragalactic star formation, which is traced by strong fine structure lines stemming from PDRs.\\  
\indent At a distance of 910 pc, the B0V star HD211880 ionizes the edge of the molecular cloud L1204, creating S140, a visible \HII\/ region and PDR \citep{Crampton1974}. 
The dense molecular cloud hosts a cluster of embedded massive young stellar objects (YSOs) only 75" from the \HII\/ region  \citep[e.g.][]{Beichman1979,Minchin1993}.
S140 and its surroundings were extensively observed in molecular tracers \citep[e.g.][]{Minchin1993,Minchin1995,vdt2000,persson2009} and the fine structure lines, \CII\/ \citep{boreiko1990,white1991,li2002}, \CI\/ \citep{minchin1994} and \OI\/ \citep{Emery1996}. An ionized component around IRS1 has been detected by \citet{hoare2006} and \citet{trinidad2007} at high resolution in the cm continuum observations. 
\citet{spaans1997} model S140 with a nearly edge on geometry, a clumpy, inhomogeneous medium and a radiation field of 150 Draine\footnote{in units of 2.7$\times 10^{-3}$ erg cm$^{-1}$ s$^{-1}$ in the FUV wavelength range
from 912~\AA\/ to 1110~\AA.} units. \\
\indent Two bi-polar outflows can be found in the S140 cloud \citep[e.g.][]{Weigelt2002,Preibisch2002}. One is traced through observations of bow-shock like features
and H$_2$ knots, indicating hot, shocked gas, lined up northeast of
IRS1 over several degrees, while the other, perpendicular to it, has been observed in CO.\\
\indent In this paper, we describe and model the \CII\/, high-$J$ CO and HCO$^+$ observations in the S140 region, observed as part of the {\it Warm and dense ISM} key project (WADI, Ossenkopf et al., KPGT$\_$vossenko$\_$1) using the HIFI \citep{de Graauw2010} instrument on the Herschel Space Observatory \citep{Pilbratt2010}.  The goal of these studies it to investigate and compare the physical properties of the different gas phases in two regions of the S140 cloud, the star forming region IRS1 and the ionization front (IF from now on).

\section{Observations \& Analysis}

\subsection{HIFI observations}

For this paper, two single pointing, frequency switching spectra in bands 1a, 4a and 4b from the Priority Science Phase and a \CII\/ OTF map from the Science Demonstration Phase were analyzed.
The single pointing data were taken towards S140-IRS1 at RA=22h19m18.21s, Dec=63$^\circ$18$'$46.9$''$ (J2000) and  towards a position in the IF at  RA=22h19m11.53s, Dec=63$^\circ$17$'$46.9$''$ (J2000).  The fully sampled \CII\/ map is centered on RA=22h19m18.88s, Dec=63$^\circ$18$'$52.9$''$ (J2000), has a length of $206"$ and crosses the PDR region from the IF through IRS1 into the dense molecular cloud (Fig. \ref{cut_cii}, right). The data presented here were taken with the wide band spectrometer (WBS) at 1.1 MHz resolution. For the FSW data, the upper and lower sidebands were processed separately. All observational parameters are summarized in Table~\ref{tab_observation}.
The data reduction was done using HIPE 3.0 \citep{hipe}. A forward efficiency of 96\%\/ and main beam efficiencies of 0.72, 0.74, 0.76 and 0.65 at 500, 950, 1112, 1900~GHz, respectively (R. Moreno, priv. comm.), were used to convert the data to main beam temperatures $T_{\rm mb}$. 

\begin{table*}
\caption{Summary of the HIFI observations used and basic results}
\scriptsize
\label{tab_observation}
\begin{center}
\resizebox{18.5cm}{!}{
\begin{tabular}{lrcl|rrrrr|rrrrr}
\hline
\multicolumn{1}{c}{Transition} &
\multicolumn{1}{c}{$\nu_{\rm line}$\hspace*{0.0cm}} &
\multicolumn{1}{c}{\hspace*{-0.1cm}HPBW }&
\multicolumn{1}{c|}{Mode$^1$} &
\multicolumn{5}{c|}{S140-IRS1}&
\multicolumn{5}{c}{S140-IF}\\
\cline{5-9}
\cline{10-14}
&&&&$t_{\rm source}$ & rms$^2$ &   $\int Tdv$\hspace*{0.2cm}& $v_{\rm lsr}$ \hspace*{0.2cm}&
$\Delta v$\hspace*{0.3cm} &$t_{\rm source}$ & rms$^2$ &   $\int Tdv$ \hspace*{0.2cm}&
$v_{\rm lsr}$ \hspace*{0.2cm}&  $\Delta v$\hspace*{0.3cm}  \\
& $[$GHz$ $]\hspace*{0.0cm} & $''$ & & $[$s$]$ & $[$K$]$ & (K~km s$^{-1}$) & (km s$^{-1}$) & (km s$^{-1}$) & $[$s$]$ & $[$K$]$ & (K~km s$^{-1}$) & (km s$^{-1}$) & (km s$^{-1}$) \\
\hline
CII{} & 1900.537 & 12 &  OTF & 12 & 0.9  &  59.63 (0.99) &  -7.26 (0.05)
& 6.82 (0.15) & 12 &0.9 & 96.91 (0.60) &  -8.00 (9.2e-03) & 3.05 (0.02)  \\
$^{13}$CO(5--4) & 550.926  & 43 & FSW & 480 & 0.02 & 85.74 (2.48) &  -7.24 (0.06) & 4.02 (0.14) & 240 & 0.03 & 4.56 (0.04) & -7.89 (0.01) & 2.85 (0.03)  \\
C$^{18}$O(5--4) & 548.831  & 43 & FSW & 480 & 0.02 & 17.29 (0.07)  & -7.24 (5.7e-03) & 2.98 (0.01) & 240 & 0.03 & 0.47 (0.03) & -7.66 (0.08) & 2.21 (0.17)  \\
HCO$^+$(6--5) & 535.062 & 43 &  FSW & 480 & 0.02 & 38.96 (4.90) &  -6.77 (0.23) &  3.75 (0.57) & 240 & 0.16 & 0.67 (0.28) &  -7.52 (0.74) &  3.48 (1.63)  \\
C$^{18}$O(9--8) & 987.56 & 23 & FSW & 66 &  0.1 & 2.60 (0.07) &  -7.03
(0.04) &2.92 (0.09) &\ldots&\ldots&\ldots&\ldots&\ldots \\
$^{13}$CO(10--9)&1101.35 & 20 & FSW & 156 &0.15 & 23.41 (0.73)& -6.99
(0.06) &  3.93 (0.15)&\ldots&\ldots&\ldots&\ldots&\ldots \\
\hline
\end{tabular} }
\end{center}
$^1$  OTF = On-The-Fly, FSW = frequency-switch, OFF position =
22h18m37.01s, 63$^\circ$14$'$17.9$''$;\hspace*{0.2cm}
$^2$ At a velocity resolution of $\Delta$v=0.7 kms$^{-1}$
\end{table*}

\subsection{Complementary data}

We have combined the HIFI data with observations
of all CO isotopes in the 2--1 and 3--2 transitions, of $^{12}$CO
in the 4--3 and 7--6 transitions, HCO$^+$ and H$^{13}$CO$^+$ in
the 3--2 and 4--3 transitions, and both fine structure lines of
atomic carbon [C{\sc I}], taken at the KOSMA 3m sub-mm telescope,
the IRAM 30 m telescope, and the JCMT.
For easy comparison of flux values, all data have been smoothed to the
$80"$ resolution of the KOSMA 345~GHZ data. The procedure
to derive appropriate scaling factors for the single-point
observations with different beams, based on the spatial
distribution measured in the SPIRE 250$\mu$m and SCUBA 450$\mu$m map, is described in
detail in the Appendix. 

\subsection{PDR modeling}

\indent To model the physical parameters in IRS1 and the IF, we fit the data by the
KOSMA-$\tau$ PDR code \citep{Roellig2006}. The PDR is represented 
by an ensemble of spherical clumps with
a power-law clump-mass spectrum \citep{cubick2008}. Details are
summarized in Appendix B.
Each clumpy PDR ensemble
has five free parameters: the average density of the clumps in the ensemble,
$n_{ens}$, the ensemble mass, $M_{ens}$, the UV field strength, $\chi$
 given in units of the Draine field, and the minimum and maximum mass
of the clump ensemble, [$M_{min}, M_{max}$].  
We fit absolute line intensities, using the available
ground-based observations and the
HIFI lines of the CO isotopes, HCO$^+$, atomic and ionized carbon. Simulated 
Annealing \citep{ingber1993} was used to find the optimum parameter combination. 
Our model ignores mutual shielding and illumination effects between
different clumps and/or ensembles, i.e. optical depth effects are only considered within individual clumps .\\

\section{Results}
\label{sect_dynamics}

\begin{figure}
\begin{center}$
\begin{array}{cc}
\includegraphics[clip,trim=10 130 20 25,angle=90.0,width=4.500cm]{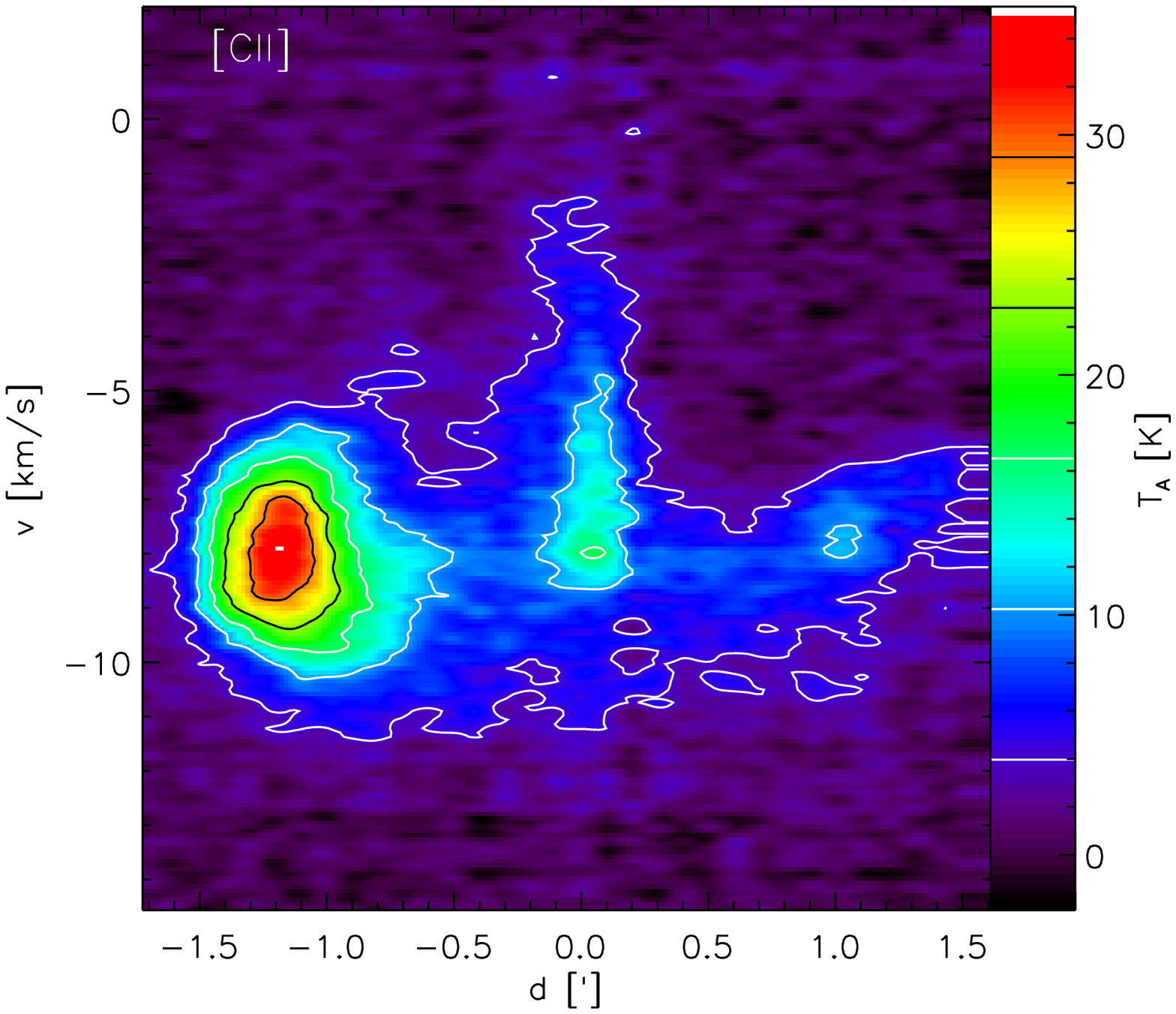} &
\includegraphics[clip, trim= 40 0 80 30  ,width=4.5cm]{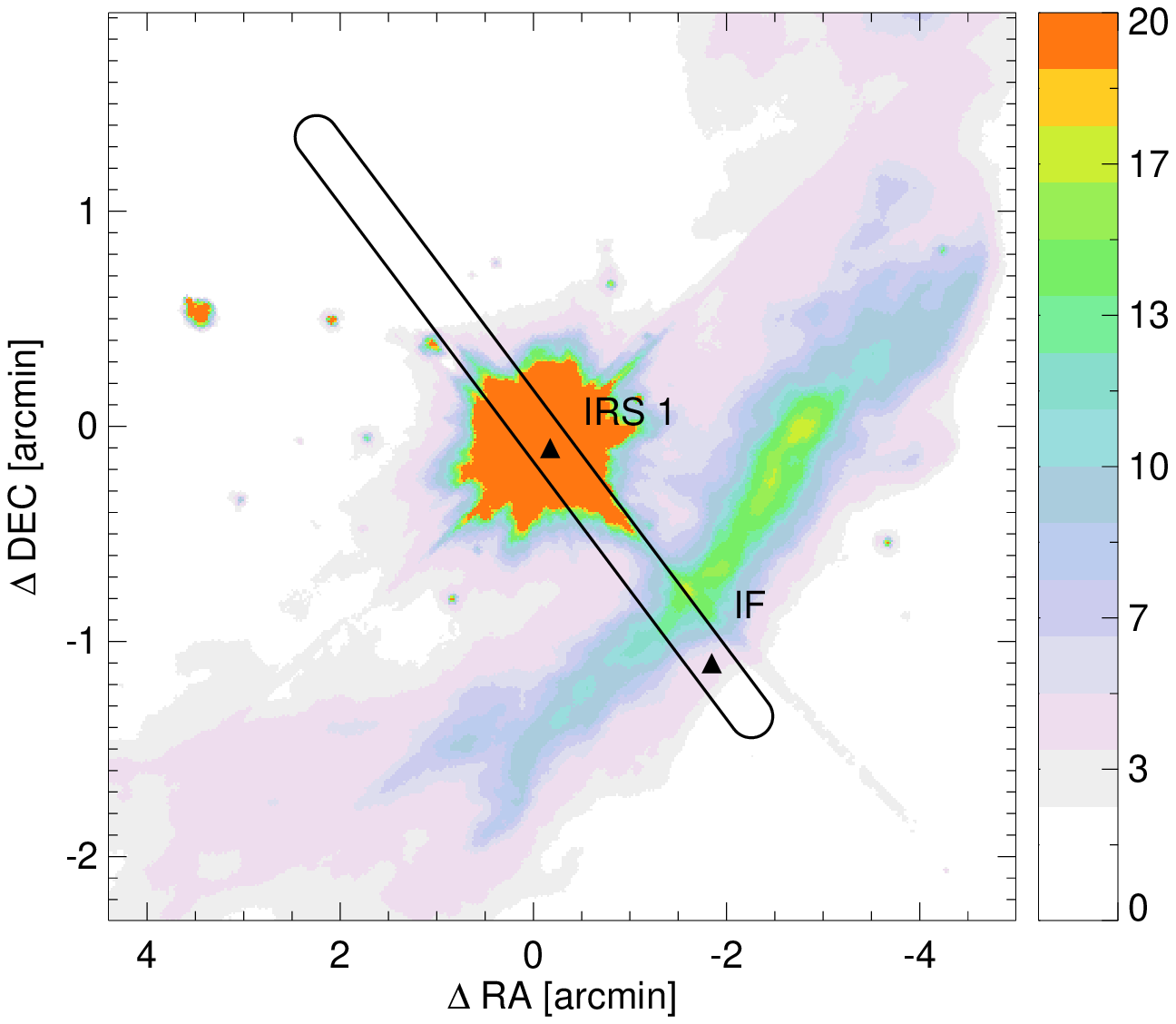} \\
\end{array}$
\end{center}
\caption{Fig. 1 Left: Position-velocity diagram of the \CII\/ emission from the IF into the cloud, centered on RA=22h19m18.88s, Dec=63$^\circ$18$'$52.9$''$ (J2000). IRS1 is at found at -7.5$"$. Right: IRAC 8$\mu$m map of S140. Shown are the two positions in which individual HIFI measurements are taken as well as the location of the \CII\/ cut. HD211880 is located to the left of the pv diagram, and to the bottom right of the cut in the right panel.}
\label{cut_cii}
\end{figure}

\begin{figure}
\centering
\includegraphics[angle=0.0,width=5.0cm]{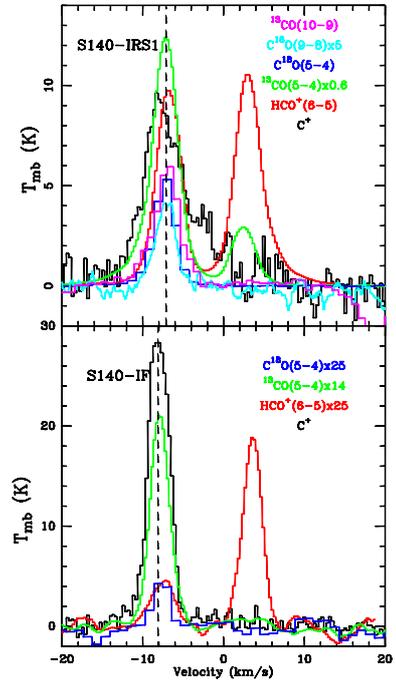}\\
\caption{HIFI spectra at IRS1 and the IF. The additional
ghost line at 4 km~s$^{-1}$ in the HCO$^{+}$(6--5) and $^{13}$CO(5--4) data is an artifact
from the FSW deconvolution of the two lines, close in IF frequency
but stemming from different receiver sidebands. Some lines are
scaled for a better comparison. The two dashed lines represent the $v_{\rm lsr}$ at $-7.1$ km~s$^{-1}$ (IRS1) and $-8$ km~s$^{-1}$ (IF).}
\label{ov_cii}
\end{figure}

\subsection{\CII\/ cut} Fig. \ref{cut_cii}, left, shows the measured position-velocity diagram of the \CII\/ cut.
We can distinguish three peaks. The first peak (IF-peak from now on), at around $-1.2'$, is located right at the
edge of the rim of the IF as seen by \citet{white1991} and represents the \CII\/ emission from the IF. A second, much broader peak is
found around $0'$, in the vicinity of IRS1 (therefore called IRS1-peak). The third peak is located at $1'$ into the dense cloud. \\
\indent At the IF position, we see symmetric \CII\/ line profiles and no
gradient in line width or centroid velocity. We find neither
indications of a pressure gradient nor of an evaporation flow
from the PDR surface.
Looking at archival IRAC 8~$\mu$m data, the peak of the \CII\/ integrated intensity at the IF-peak is
consistent with that of IRAC 8~$\mu$m, tracing the PAH and evaporating VSG emission. \\
\indent The emission at the IRS1-peak is $10"$ offset from the young stellar object IRS1 and coincident with the locations of the deeply embedded source Submm-2, which was first described by \citet{Minchin1995} and the strong H$_2$ knot \#1 found by \citet{Preibisch2002}. Compared to the K-band data of \citet{Weigelt2002}, the \CII\/ peak is close to the edge of what they describe as the wall of the outflow cavity carved into the material by the jet. The strong \CII\/ emission seen here is probably created at this
interface, either by irradiation of the outflow walls  \citep[e.g.][]{bruderer2009a,bruderer2009b} or shock interactions, as traced by the H$_2$ knots.
The \CII\/ lines are very broad
with a pronounced red wing tracing an outflow of ionized carbon, which is visible in Fig. \ref{ov_cii} between $-5$ km~s$^{-1}$ and $0$ km~s$^{-1}$. A 2-component Gaussian profile with components
at $-8$ km~s$^{-1}$ and $-5.5$ km~s$^{-1}$ fits the observed \CII\/ line
profile. Since the $-5.5$ km~s$^{-1}$ component of the \CII\/ emission
is not observed in any of the other molecular tracers, it does not
arise in the outflows, but may originate in the ionized medium.
\indent Between the two \CII\/ peaks, the \CII\/ emission drops significantly, indicating a spatial separation between the IF component and the component around IRS1. 
This drop in \CII\/ intensity between the IF-peak and IRS1-peak roughly coincides with a local minimum of the dust temperature, where neither the heating from the internal IRS1 source nor the
external star (Lester et al., 1986) are efficient.\\
\indent In the region where the third peak is seen in \CII\/, the SPIRE 250 $\mu$m map also shows a secondary peak
(Juvela et al., priv. comm.), which points to a deeply embedded source. The \CII\/ line has pronounced wings, at a slightly higher velocity than in IRS1. This might indicate a separate
outflow or material shocked by the outflow from IRS1. \\
\indent Several spectra were taken with HIFI at the IRS1 and IF position. Overlay of those spectra of different tracers shows
that with the exception of \CII\/ all the molecular lines at IRS1 have similar
line shapes (Fig. \ref{ov_cii}). The \CII\/ line shows a strong red wing, which was also not detected in the previous coarser spatial resolution
observations by \citet{boreiko1990}. The centroid velocity, $v_{\rm cen}$, for all the molecular
lines range from $-6.8$ to $-7.4$ km~s$^{-1}$ and agree well with a
source velocity of $-7.1$ km~s$^{-1}$ as derived from previous observations
 \citep[e.g.][]{vdt2000,persson2009}.  Although
somewhat fainter in intensities (particularly HCO$^+$ 6--5), several of
the molecular lines are still detected at the IF position. The line
widths at the IF position are up to a factor of two narrower compared to the widths at
the position of IRS1. At the IF position the centroid velocities fall
between $-7.5$ to $-8.0$ km~s$^{-1}$, corresponding to a slight blue shift compared to
the velocities at the IRS1 position.\\
Detailed discussion of the \CII\/ line profiles
will be taken up in a later paper by Mookerjea et al. (in
preparation).

\subsection{PDR model in IRS1}

\begin{figure}
\centering
\includegraphics[angle=0.0,width=6.5cm]{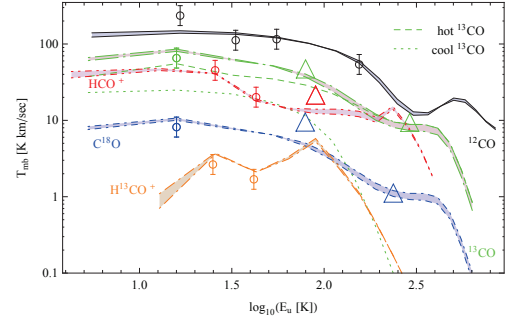}
\caption{Clumpy PDR model fit to the observed CO and HCO$^+$ line integrated intensities in IRS1, shown as function of the upper
level energy. HIFI measurements are depicted as open triangles, 
complementary data points as open circles. The range of model intensities spanned by the total mass range of 54--250 M$_\odot$ for the cool component is shown as shaded area for the $J$-lines.
}
\label{fig_modelfit}
\end{figure}

Since the PDR region arising from the irradiated outflow walls
is illuminated both from the outside by the B0V star HD211880
and from the inside by the cluster of YSOs around IRS1, we assume
a two component fit with a hot component originating at
IRS1 with strong FUV illumination, but only a small fraction of the total
mass, and a cooler component that provides the bulk of the material.
Owing to the much coarser resolution of the available hydrogen and carbon radio recombination lines  \citep{smirnov1995,wyrowski1997}  it is difficult to ascertain the fraction (if any) of \CII\/ emission being contributed by the ionized gas associated with
IRS1. Based on the Gaussian decomposition described in Sect. 3.1,
the \CII\/ intensity from the PDR must fall between the
60 K km~s$^{-1}$ of the total integrated line and the 37 K km~s$^{-1}$
of the $-8$ km~s$^{-1}$ component only. Consequently, we obtain two
different limiting fits that differ mainly by the mass of
the cool component falling between 54 and 250 M$_\odot$.
The best fit result is shown in Fig. \ref{fig_modelfit}. The corresponding model
parameters for both components can be found in Table \ref{param}.  Figure \ref{fig_modelfit} also indicates the contributions of the two components
to the emission. 

\begin{table}
\begin{center}
\caption{\label{param} The parameters for the two ensembles in the PDR model in IRS1.}
\resizebox{9cm}{!}{
\scriptsize
\begin{tabular}{lclccccc }
\hline
$\langle n_{\rm ens}\rangle$& M$_{\rm ens}$& $\chi$ & [$M_\mathrm{min},M_\mathrm{max}$] &{\bf N$^{1)}$\hspace*{0.15cm} }& $f_{\rm V}$&  $ f_{A}^{2)}$ \\
(cm$^{-3}$) & ($M_\odot$) &(Draine) & ($M_\odot$) & &  &  \\
\hline
{\bf Hot:} & & & & & &  \\
1.8~$10^6$ &  14 & 2.3\,$10^5$ & [0.008,7] & 151 & {\bf  0.42 -- 0.73} &1.3 \\
{\bf Cool:} & & & & & &  \\
1.3~$10^6$ & 54--250 & 27 & [0.008,27] &  413 - 1913 & {\bf 0.07 -- 0.34} &  0.6  - 2.6 \\
\hline       
\end{tabular}  
}
\end{center}
$^{1)}$ Number of all clumps; $^{2)}$ assuming hot clumps are situated approximately 7$"$ away from IRS1;   $f_{\rm V}$ volume filling factor within a shell with $R_{\rm V,inner}$=5$"$ and $R_{\rm V,outer}$=11--13$"$; $f_{A}$ area filling factor with $R_{\rm A}$=7$"$ for the hot component. For the cold component,  $R_{\rm V,inner}$=11-13$"$, $R_{\rm V,outer}$=40$"$ and $R_{\rm A}=40"$ 
\end{table}

\begin{table}
\begin{center}
\caption{\label{model_fs} PDR model fit to the observed atomic fine structure line integrated intensities in IRS1. Shown are the ranges spanned by the two fits.} 
\begin{tabular}{lcccc}
\hline
species & total &hot &cool & obs. \\
 & (K km~s$^{-1}$) & (K km~s$^{-1}$) & (K km~s$^{-1}$) & (K km~s$^{-1}$) \\
 \hline
\CII &19.3 -- 21.9 & 16.5  & 2.8 -- 5.4  &  37.\\
\CI(1--0) & 9.7 -- 11.0 & 1.5  & 8.2 -- 9.5 &  11.4 \\
\CI(2--1) & 6.9 -- 8.5  &  1.7  & 5.2 -- 6.8 & 20.5 \\
\hline		
\end{tabular}
\end{center}
\end{table}

The reduced $\chi^2$  of the fit is 2.4. We applied a generic 30\% error
to all data points. The model slightly underestimates the mid-
$J$ lines of the CO isotopologues and of HCO$^+$. This is the energetic
region where neither the cool nor the hot component are very
bright, so that this may hint towards a third intermediate
component. However, the available data do not allow to reliably
derive parameters of a third component. To see the modeled FUV flux of the hot component, the clumps have to be at a distance of 7$"$ from IRS1. 
At this distance, an area filling factor on a spherical shell of 1.3 is obtained,  i.e. the hot component shields the cool ensemble from most of the FUV flux. 
To obtain reasonable values of volume filling factors, i.e. $0.42 \leq f_{\rm V} \leq 0.73$, the inner and outer radii of the shell have to be roughly equivalent to 5$"$ and 11--13$"$, respectively. While at the larger radii, the FUV flux seen by the hot component would be somewhat lower than predicted by our model, this discrepancy is still small compared to the uncertainties in the line estimates and thus the model parameters. 
The model parameters of the cool ensemble are particularly
interesting. The FUV field is lower than expected from the
externally illuminating star HD211880. This suggests that the
prominent edge-on PDR at the IF blocks the FUV radiation.

\subsection{PDR model in the IF}

For the IF position we are not yet able to provide a full model
fit to the data because of the lower numbers of detected lines. We can,
however, derive some qualitative conclusions.
The mid- and high-$J$ lines of CO and HCO$^+$ can only be
explained by a hot gas component with densities of the order of
10$^6$ cm$^{-3}$ and a local FUV intensity in the order of
10$^5$ Draine units. The external radiation from HD 211880 cannot provide more than a few thousand Draine units. An additional,
so far unknown UV source seems to be present.
The total gas mass in the beam at the
IF is on the order of 10~$M_\odot$.

\section{Discussion and Conclusions}

About 90\% of the \CII\/ emission around IRS1 stems from the hot component
near the embedded IR sources. A different heating mechanism accounts
for the [CII] from the emission in the ionizing front. This is reflected in our choice of two separate models for the two regions. \\
\indent Previous models of PDR emission in S140 \citep{spaans1997} describe the S140 region as an inhomogeneous medium with $\langle n_{\rm H} \rangle$  = $7 \times 10^3$cm$^{-3}$, a filling factor of 30\% and clumps of 0.04~pc size. The radiation field, $\chi$, of 150 Draine units they derive is much lower than the one derived in this work, however, these authors include the IRS1 cluster as an embedded heating source in their model, yet do not distinguish between the two regions of \CII\/ emission - the IRS1 region and the IF.
Moreover, they could not include high-$J$ CO lines,
which are important tracers of the hot gas. 
 In our two component approach, the high-$J$ CO lines on one hand tell us how strong the hot ensemble has to be illuminated by the FUV and they determine how the total
mass is distributed between hot and cool component.
High-$J$ CO lines will be especially valuable since they, in the context of PDR models, can only be produced by the very high FUV fields while lower-$J$ lines have always more mixed origins and hence pose weaker constraints for the multi component model. With the exception of the high-$J$
$^{13}$CO(10--9) and C$^{18}$O(9--8) lines most other data points can be
explained with much more moderate FUV fields.\\
\indent When repeating our calculations using only pre-HIFI data
we reproduce the parameters obtained in previous models \citep[e.g.][]{spaans1997,koester1998}. It
becomes apparent that only HIFI data
can reveal additional components deep inside the cloud.
 The low mass of the hot component together with the low FUV field
of the cool component suggests that only a small fraction of the
material sees the full FUV field, supposedly where the outflows created
cavity walls at the edge of the ambient material. The bulk of the material is effectively shielded from the FUV photons.\\
\indent At the IF, first qualitative results hint at a so far unknown heating source and a two component density structure. A full PDR-model on this position (R\"ollig et al., in preparation) will reveal further properties of the material at the IF. \\
\indent HIFI's high spectral resolution observations allow us to identify the general velocity fields of the molecular and ionized gas around IRS1 and trace the dynamics of the region, such as the blue outflow emission associated with the SPIRE-peak. This is possibly due to another, still deeply embedded object, which is driving its own outflow.

\begin{acknowledgements}
HIFI has been designed and built by a consortium of institutes and university departments from across
Europe, Canada and the United States under the leadership of SRON Netherlands Institute for Space
Research, Groningen, The Netherlands and with major contributions from Germany, France and the US.
Consortium members are: Canada: CSA, U.Waterloo; France: CESR, LAB, LERMA, IRAM; Germany:
KOSMA, MPIfR, MPS; Ireland, NUI Maynooth; Italy: ASI, IFSI-INAF, Osservatorio Astrofisico di Arcetri-
INAF; Netherlands: SRON, TUD; Poland: CAMK, CBK; Spain: Observatorio Astronómico Nacional (IGN),
Centro de Astrobiología (CSIC-INTA). Sweden: Chalmers University of Technology - MC2, RSS \& GARD;
Onsala Space Observatory; Swedish National Space Board, Stockholm University - Stockholm Observatory;
Switzerland: ETH Zurich, FHNW; USA: Caltech, JPL, NHSC.\\
The work on star formation at ETH Zurich is partially funded by the Swiss National Science Foundation (grant nr. 200020-113556). 
This program is made possible thanks to the Swiss HIFI guaranteed time program.\\
 This work was supported by the German
 \emph{Deut\-sche For\-schungs\-ge\-mein\-schaft, DFG\/} project
  number Os~177/1--1. 
We thank the members of the Herschel key project "Galactic Cold
Cores: A Herschel survey of the source populations revealed by
Planck" lead by M. Juvela (KPOT$\_$mjuvela$\_1$) for providing us with
the results of the SPIRE 250 $\mu$m mapping and fruitful discussions.\\
We would also like to acknowledge the use of  the JCMT CO(3--2) archival data (PI M. Thompson, M08BU15).     
 A portion of this research was performed at the Jet Propulsion Laboratory, California Institute of Technology, under contract with the National Aeronautics and Space administration.
 We would like to thank an anonymous referee for constructive comments.
\end{acknowledgements}

\Online

\begin{appendix}
\section{Comparison of line profiles taken with different beams }

The different lines discussed here were measured at various
telescopes and at different frequencies so that they all
represent somewhat different spatial resolutions. To allow
a comparison in terms of a physical interpretation, they have
to be translated to a common resolution, so that they stem from
the same area on the sky. As a reference, we use $80"$, the
resolution of the KOSMA observations in the 3--2 transition of
the CO isotopes. In principle, all data taken at a finer resolution
can be resampled to this beam if the mapped area is large enough.

Unfortunately, most HIFI observations were only single-point
observations, not full maps, so that such a convolution is
impossible. To derive scaling factors that describe the
translation between the measured intensities and the intensity
that would be obtained in an $80"$ beam, we have to assume a
source geometry of the emission. Instead of using any analytic
geometry, we use the actually measured distribution of warm dust
seen in the sub-mm. By assuming that the spatial distribution of
all PDR tracers roughly follows the warm dust we derive scaling
factors for the line intensities at different beam widths, by
convolving the sub-mm continuum map with the different
beam sizes and picking ratios between the convolved intensities
at the measured positions.\\
\indent As a continuum map at IRS1, we used the combination of the SPIRE 250\micron\ 
($18"$ beam size\footnote{$http://herschel.esac.esa.int/Docs/SPIRE/html/spire\_om.html$} ) and SCUBA 450\micron\ image ($8"$
beam size, Holland et al. 1999), because IRS1 is saturated in SPIRE
250\micron, whereas the observed area of SCUBA 450\micron\ is too small
to make a convolution map with the beam size of $80"$. We regrid
the SCUBA 450$\mu$m image to the same grid as SPIRE 250 $\mu$m, determine
the scaling factor between the SCUBA 450$\mu$m and the SPIRE 250 $\mu$m
maps from the overlapping area, and replace the saturated pixels
of SPIRE 250$\mu$m with the scaled SCUBA data.
As this combination implies some arbitrariness, we have tested
four different approaches. We either derive the scaling factor
in a least-square fit using (1) all the valid overlapping pixels
or (2) only overlapping pixels with SPIRE 250$\mu$m
$> 100$ Jy/beam. When replacing
SPIRE pixels, we either replace only saturated pixels (1),
or replace the full square area (5 $\times$ 5 pixels, 2) containing
all saturated  pixels. The combination of these provide 4
different beam scaling factors, which are consistent within
3\%. Taking the average of these 4 values, we derive the final factors
as shown in Table \ref{scl}.
A direct convolution to $80''$ was possible for the ground-based
maps that were observed with a smaller beam, such as the JCMT
CO(3--2) map. All the resulting intensities are summarized in Table \ref{tab_groundbased}.

\begin{table}[h]
\begin{center}
\caption{\label{scl} The scaling factors between the different beam size observations, which should be multiplied to the line intensity to estimate the one in a $80"$ beam.}
\begin{tabular}{lc}
\hline
IRS1 &  \\
beam size ($"$) & scaling factor  \\
19.7 &  0.40    \\
42.0  & 0.57  \\
43.0  &  0.55 \\
55.0  & 0.70  \\
80.0  & 1.00  \\
		  
\hline		
\end{tabular}
\end{center}
\end{table}

\begin{figure}
\centering
\includegraphics[angle=0.0,width=\columnwidth]{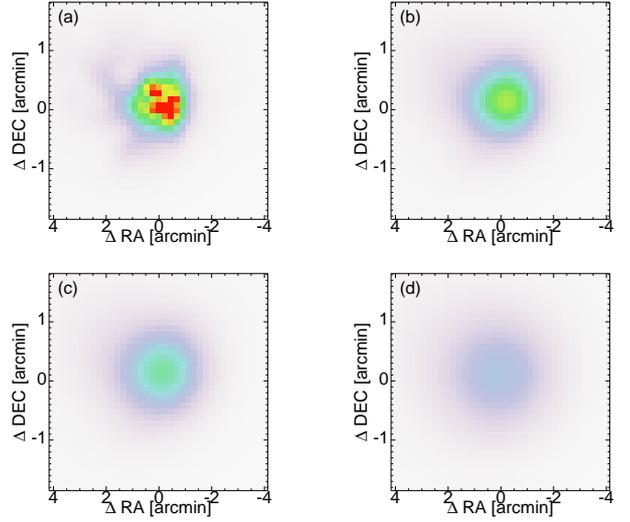}
\caption{a) 
The spatial distribution of warm dust obtained by combining
the SPIRE 250$\mu$m map with the SCUBA 450$\mu$m
map (see text). The HIFI beam is assumed to $"$see$"$ convolved images
at beam sizes of (b) $42"$, (c) $55"$, and (d) $80"$.
The color scale is common for all the figures. The coordinates are relative to the position of IRS1. Thus the flux ratios at (0,0) in these images are the values in Table \ref{scl}.}
\label{fig_app}
\end{figure}

\begin{table}
\caption{List of the complementary data. All fluxes have been scaled to $80"$ beam size. Values in brackets were not included in the fit. }\label{tab_groundbased}
\begin{center}
\begin{tabular}{cccc}
\hline
tracer & IRS1  & beam & telescope\\
 & (K kms$^{-1}$) & (") &  \\
\hline  
 CO(2--1) &  (130)  & 130 & KOSMA$^a$ \\
 $^{13}$CO(2--1) &  (36.1) & 130 & KOSMA$^a$  \\
 C$^{18}$O(2--1) &8.2  & 11.2 & IRAM$^b$ \\
 CO(3--2) &  112.2  & 11.4 & JCMT$^c$  \\
 $^{13}$CO(3--2) & 75.5  & 80 & KOSMA$^a$  \\
 CO(4--3) & 115.5 &  57 & KOSMA$^d$ \\
 CO(7--6) & 54.0  & 42 & KOSMA$^d$  \\
 CI($^3P_1 - ^3P_0$) & 11.4 & 55 & KOSMA$^d$  \\
CI($^3P_2-^3P_1$) & 20.5  & 42 & KOSMA$^d$ \\
HCO$^+$(3--2) & 45.2  & 19.7 & JCMT$^e$ \\
H$^{13}$CO$^+$(3--2) & (1.46) &  130 & KOSMA$^f$  \\
HCO$^+$(4--3) & 20.12  & 13.2 & JCMT$^e$   \\
H$^{13}$CO$^+$(4--3) & 1.69 &  80 & KOSMA$^f$ \\
\hline
\end{tabular}
\end{center}
$^a$: D. Bremaud, Diplomarbeit, ETH, $^b$ Kramer et al. 1998, A\&A, 329, 249 , $^c$: M. Thompson, JCMT archive, $^d$: S. Heyminck, priv. comm., $^e$: S. Bruderer, priv. comm., $^f$: E. Buenzli, Diplomarbeit ETH

\end{table}

\section{The clumpy PDR model}

To model the far-infrared line emission from S140 we used a superposition
of spherical clumps described by the KOSMA-$\tau$ PDR model \citep{Roellig2006}
that represent an ensemble of clumps with a fixed size-spectrum \citep{cubick2008}.
The KOSMA-$\tau$ PDR model simulates a spherical cloud with a radial 
density profile given by 
\begin{equation}
n(r) = n\sub{s} \begin{cases}
 c^\delta & \text{ for}~ r \le R/c\\
 (r/R)^{-\delta}&  \text{ for}~ R/c<  r \le R\\
 0 & \text{ for}~ r > R\;.\\
\end{cases}
\end{equation}
The constant $c$ determines the dynamic range that is covered by the
power-law density decay.
The spectrum of PDR clumps is characterized by the clump mass spectrum
\begin{equation}
{d N \over d M} = a M^{-\alpha}\;,
\end{equation}
where the factor $a$ is determined by the total mass of clumps
within the beam, $M\sub{ens}$, and the mass-size relation
\begin{equation}
M = C R^{\gamma}\;,
\end{equation}
which implicitly defines the surface density of the individual clumps
$n\sub{s}$. The factor $C$ is determined by the average ensemble
density $n\sub{ens}$. Guided by the clump-decomposition results
from \citet{Heithausen} we
fix the parameters of the spectrum to $\alpha=1.8$, $\gamma=2.3$,
$\delta=1.5$, and $c=5$. The boundaries of the clump size distribution
$M\sub{min}$ and $M\sub{max}$ are used as free parameters with the
constraint that the maximum clump mass cannot exceed half of the
mass of the total ensemble.

Every individual clump is treated as a spherically symmetric configuration
illuminated by an isotropic external UV radiation field, specified in terms
of the average interstellar radiation field, $\chi_0=2.7\times 
10^{-3}$~erg~cm$^{-2}$~s$^{-1}$ in Draine units, and cosmic rays
producing an average ionization rate, 
$\zeta\sub{CR}= 5 \times 10^{-17}$~s$^{-1}$. The internal velocity
dispersion of molecules within the clumps is fixed to 1~km~s$^{-1}$.
The model computes the stationary chemical and temperature
structure by solving the coupled detailed balance of heating,
line and continuum cooling and the chemical network using the
UMIST data base of reaction rates \citep{UMIST} expanded by separate
entries for the $^{13}$C chemistry \citep[see][for details]{Roellig2007}. 
The chemical network currently does not include $^{18}$O, so that
C$^{18}$O predictions can only be obtained by scaling the $^{13}$CO
values ignoring fractionation between the two species. As this ignores the different 
self-shielding of $^{13}$CO and C$^{18}$O, the model results for C$^{18}$O
are less reliable than for the other lines.

In the superposition of clumps, the line emission from the different
clumps is simply added assuming that the velocity dispersion between 
the clumps is large enough, so that they do not shield each other in
position-velocity space. This is valid for most species, except for the \OI\/
emission which is optically very thick ($\tau \ga 100$), so that
the lines are much broader than the velocity distribution. Therefore,
the model is not able to give any reliable estimate for the
\OI\/ intensities. For the continuum extinction of the UV
radiation, the situation is different. There, mutual shading of the 
clouds is relevant, leading to the concept of different clump
ensembles that "see" different UV fields if the average ensemble
extinction exceeds an $A\sub{V}$ of about unity.

The parameters of the clump mass spectrum imply that most of the
mass is actually contained in the largest clumps that also have
the maximum column density or $A\sub{V}$, respectively. The dependence of  $A\sub{V}$ and clump size on clump mass is given by

\begin{equation}
R = 5.3 \times 10^{18} \left( \frac{M [M_{\odot}] }{n [cm^{-3}]} \right)^{\frac{1}{3}}
\end{equation}

and 

\begin{equation}
A\sub{V} =1.6018 \times 10^{21} n[cm^{-3}] R[cm]
\end{equation}

 In terms of
the total clump surface or the total solid angle of the different
clumps we find about equal contributions from each logarithmic
mass bin in the ensemble \citep[Eq. 16 in][]{cubick2008}. Consequently,
we find a non-trivial dependence of the intensity in the different HIFI
lines on the clump mass. Figs. 3 and 5 from \citet{cubick2008} show
that the [C{\sc II}] emission is dominated by the largest clumps
while the high-$J$ CO lines are dominated by the smallest clumps.
A complex, non-monotonic behavior is observed for the lines from
atomic oxygen and mid- to low-$J$ CO isotopes.

As the clump spectrum is purely observationally based, it has no
direct relation to a stability criterion. In fact we find that
both the most massive clumps are unstable to gravitational collapse
and the smallest clumps will be dispersed on the timescale of a
few million years. Therefore, the spectrum can only be considered
as a snapshot of interstellar turbulence that reflects the density
structure over a timescale of $10^6-10^7$ years. The assumption of
a steady-state chemistry and energy-balance is therefore only 
applicable if all rates are faster. This holds for the dense clumps
in the S140 model fit with densities above $10^4$~cm$^{-3}$, but
for lower densities, an explicitly time dependent modeling would
be required \citep{viti2006}.

\end{appendix}

\end{document}